\documentclass[letter]{aa}
\usepackage{graphics}
\usepackage{natbib}
\bibpunct{(}{)}{;}{a}{}{,}
\usepackage{amsmath}
\usepackage{upgreek}
\usepackage{color}
\usepackage{xspace}
\usepackage{hyperref}
\hypersetup{
    pdfborder={0 0 0 0},
    colorlinks=true,
    linkcolor=blue,
    urlcolor=blue,
    citecolor=blue
}

\newcommand{\uJyb}{\mathrm{\upmu Jy\ beam^{-1}}\xspace}

\def\ba{B0525$+$21\xspace}
\def\bb{B2045$-$16\xspace}

\begin{document}

\title{Probing the origin of the off-pulse emission from the pulsars B0525+21 and B2045$-$16}
\titlerunning{On the off-pulse emission from B0525+21 and B2045$-$16}

\author{B.~Marcote\inst{\ref{inst1}}\thanks{\email{marcote@jive.eu}}
    \and Y.~Maan\inst{\ref{inst2}}\thanks{\email{maan@astron.nl}}
    \and Z.~Paragi\inst{\ref{inst1}}
    \and A.~Keimpema\inst{\ref{inst1}}
}

\authorrunning{Marcote et al.}

\institute{
    Joint Institute for VLBI ERIC, Oude Hoogeveensedijk 4, 7991~PD Dwingeloo, The Netherlands\label{inst1}
\and
    ASTRON, the Netherlands Institute for Radio Astronomy, Oude Hoogeveensedijk 4, 7991~PD Dwingeloo, The Netherlands\label{inst2}
}

\date{Received ... ; accepted ...}

\abstract{Pulsars typically exhibit radio emission in the form of narrow pulses originated from confined regions of their magnetospheres. A potential presence of magnetospherically originated emission outside this region, the so-called off-pulse emission, would challenge the existing theories. Detection of significant off-pulse emission has been reported so far from only two pulsars, \ba and \bb, at 325 and 610~MHz. However, the nature of this newly uncovered off-pulse emission remains unclear. 
To probe its origin we conducted very high resolution radio observations of \ba and \bb with the European VLBI Network (EVN) at 1.39~GHz. Whereas the pulsed emission is detected at a level consistent with previous observations, we report absence of any off-pulse emission above  $42$ and $96\ \mathrm{\upmu Jy\ beam^{-1}}$ (three times the rms noise levels) for \ba and \bb, respectively. Our stringent upper limits imply the off-pulse emission to be less than $0.4$ and $0.3\%$ of the period-averaged pulsed flux density, i.e.,\ much fainter than the previously suggested values of $1$--$10\%$. Since the EVN data are most sensitive to extremely compact angular scales, our results suggest a non-magnetospheric origin for the previously reported off-pulse emission. Presence of extended emission that is resolved out on these milliarcsecond scales still remains plausible. In this case, we constrain the emission to arise from structures with sizes of $\sim (0.61$--$19) \times 10^3\ \mathrm{au}$ for \ba and $\sim (0.48$--$8.3) \times 10^3\ \mathrm{au}$ for \bb. These constraints might indicate that the two pulsars are accompanied by compact bow-shock pulsar wind nebulae.
Future observations probing intermediate angular scales ($\sim 0.1$--$5\ \mathrm{arcsec}$) will help in clarifying the actual origin of the off-pulse emission.

}

\keywords{pulsars: individual: B0525+21  --  pulsars: individual: B2045-16 -- Radiation mechanisms: non-thermal -- Radio continuum: stars -- Techniques: high angular resolution -- Techniques: interferometric}

\maketitle

\section{Introduction}

Pulsars are highly magnetized fast rotating neutron stars. The \emph{pulsed} emission from pulsars is believed to originate from highly relativistic particles streaming along the magnetic field lines in narrow regions around the magnetic axes. The radius of the light cylinder, i.e., the equatorial distance from the neutron star beyond which co-rotation speed is more than the speed of light, provides a characteristic extent of the pulsar magnetosphere. The pulsed emission originates relatively close to the neutron star surface (well within the pulsar magnetosphere) along the magnetic field lines which do not close within the light cylinder, called the open field lines.

A potential presence of \emph{unpulsed} or \emph{off-pulse} emission component has been speculated for a long time. Such a continuous radio emission present at all rotation phases of the star would remain undetectable in the typical time-domain studies of pulsars and often require interferometric probes. A presence of a pulsar wind nebula (PWN) or any emission from the supernova remnant material in the vicinity of the pulsar may contribute to such continuous emission. However, detection of off-pulse emission originated within the pulsar magnetosphere would have important implications for the pulsar emission theories.

There have been several early searches for the off-pulse emission using different techniques, e.g., \citet[][]{schonhardt1973,cohen1983,perry1985,strom1990}. Some of these studies claimed detection of off-pulse emission from a few pulsars. However, the detections were argued to have been confused with the associated PWNe or unrelated sources nearby in sky \citep[e.g.][]{hankins1993}.
More recently, \citet{basu2011,basu2012} reported more convincing detections of off-pulse emission from two pulsars, \ba and \bb, using observations with the Giant Metrewavelength Radio Telescope (GMRT) at 325 and 610~MHz. These observations were ``gated'' in off-line processing to cleanly separate the emission within the pulsed and the off-pulse rotation phases of the pulsars. The authors reported $325$-$\mathrm{MHz}$ off-pulse flux densities of $3.9 \pm 0.5~\mathrm{mJy}$ and $4.3 \pm 1.1~\mathrm{mJy}$ from \ba and \bb, respectively, which correspond to $\sim 10$ and $\sim 1\%$ of the respective pulsed flux densities averaged over the period. Based on the spatial resolution of their observations, the authors suggested the off-pulse emitting regions to be $\lesssim 0.09$ and $0.04~\mathrm{pc}$ (or $\lesssim 20 \times 10^3$ and $8 \times 10^3\ \mathrm{au}$) for \ba and \bb, respectively.

\citet{basu2011} concluded that the off-pulse emission should have a magnetospheric origin, and a physical scenario where the emission could be produced by cyclotron resonance instabilities near the light cylinder was also proposed \citep{basu2013}. Given that \ba and \bb are reasonably old pulsars, presence of static PWNe around these objects seems unlikely. However, the possibility of these low-energy pulsars driving very compact bow shock PWNe still cannot be ruled out. Moreover, \citet{Blandford73} proposed that once the surrounding supernova remnant is dissipated, the relativistic particles from the old pulsar may interact with the ISM magnetic field and radiate, creating `ghost remnants'. Very high angular resolution observations, such as those using very long baseline interferometry (VLBI), would help in disentangling the emission from surrounding material and the off-pulse emission originated within the pulsar magnetosphere, if any.
Similar characteristics of the on- and off-pulse emission, namely the scintillation effects and spectral indices, also motivated \citet{basu2012} to propose the magnetospheric origin.

With the above motivations of unveiling the actual origin of the off-pulse emission, we present our probes on milliarcsecond angular scales using European VLBI Network (EVN) observations of \ba and \bb. The EVN correlator facilitates a mode to provide visibilities binned across the pulsar period, enabling the much desired phase-resolved probes of the off-pulse emission with high resolution. Section~\ref{sec:obs} describes the observations and the data reduction. Section~\ref{sec:results} presents the obtained results. We discuss the detailed implications of these results in Sect.~\ref{sec:discussion}, followed by the main conclusions in Section~\ref{sec:conclusions}.

\section{Observations and data reduction}\label{sec:obs}

We observed each of the two pulsars, \ba and \bb, at two different epochs with the EVN. A total of 13 stations participated during these observations: Effelsberg, the Lovell Telescope, Jodrell Bank Mark2, Medicina, Onsala, Toru\'n, Westerbork single dish, Urumqi, Svetloe, Badary, Zelenchukskaya, Tianma, and Robledo. The data were recorded with a total bandwidth of 128~MHz centered at 1.39~GHz, and divided into eight subbands of 32 channels each, with full circular polarization, during correlation. Single-dish data from Effelsberg were simultaneously recorded at the station using the PSRIX backend recorder \citep{lazarus2016}. The data were online coherent dedispersed and folded in 10-s sub-integrations using 1024 bins across the respective pulsar periods, for each of the 256 frequency channels. 
These single dish data were used to provide updated ephemerides of the pulsars for the binning-mode correlation of the interferometric data.

\ba was observed on 24 February and 27 May 2017 for 7.5~h each. 3C~84 was used as fringe finder, and J0521$+$2112 as phase calibrator in a phase-referencing cycle of 3.5~min on the target and 2~min on the calibrator.
\bb was observed on 25 February and 30 May 2017 for 5.5~h each. In this case, J2148+0657 was used as fringe finder, and J2047$-$1639 as phase calibrator in a similar phase-referencing cycle of 3.5~min on the target and 2~min on the calibrator.

For each pulsar we correlated the data from all stations using the SFXC software correlator in a pulsar-binning mode \citep[see][for further details]{kettenis2014,keimpema2015}, dividing the rotation period of each pulsar ($\sim 3.75$ and $1.96\ \mathrm{s}$ for \ba and \bb, respectively) into 64 bins, with a time integration of $4$ and $2\ \mathrm{s}$, respectively. We note that the expected smearing due to interstellar dispersive effects at our observing frequency is much smaller than one bin width for either of the pulsars.

\begin{figure}[!t]
    \includegraphics[width=0.49\textwidth]{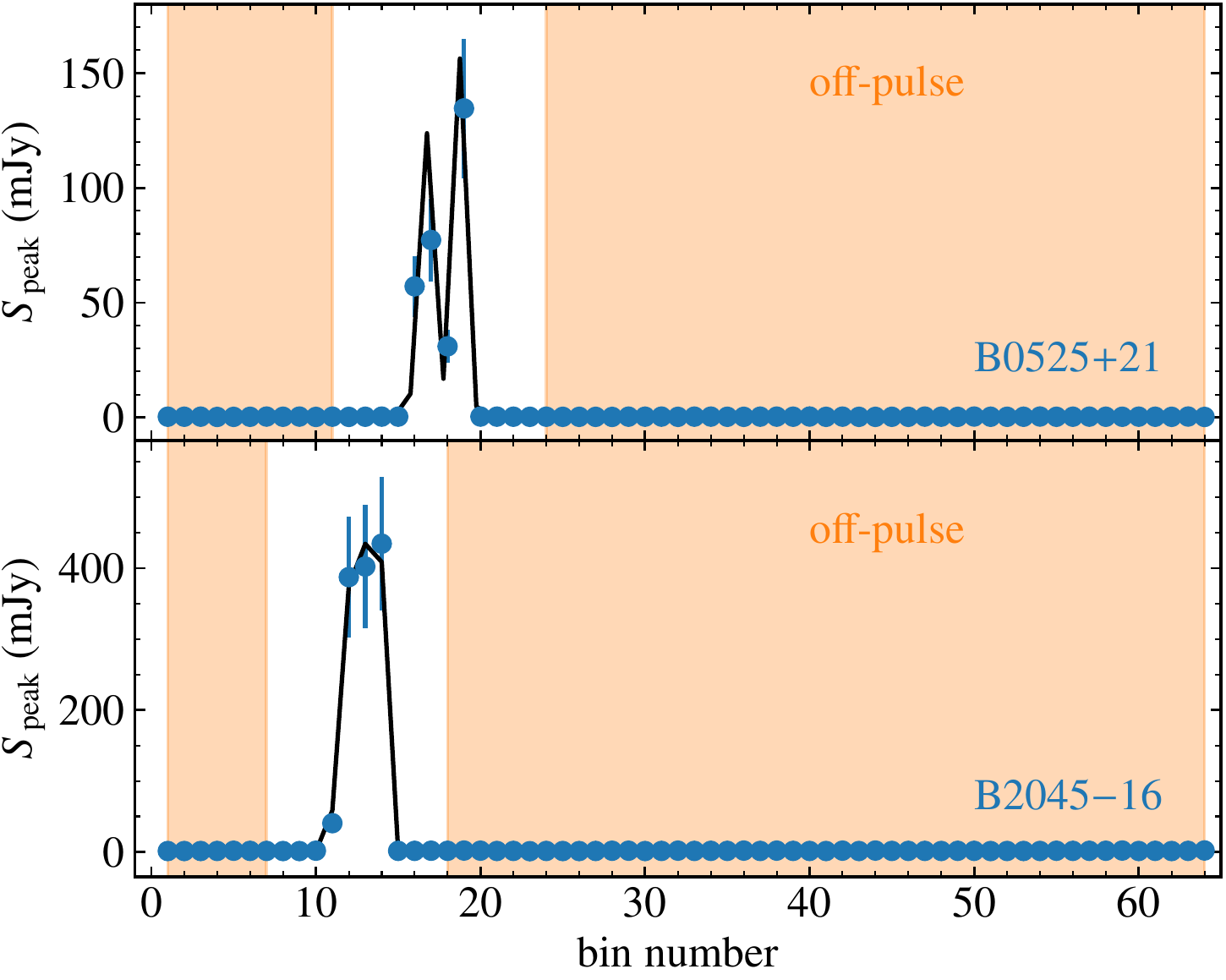}
    \caption{Light-curves folded over the rotation periods of \ba (top) and \bb (bottom). Pulsed emission is detected in bins 16--19 and 11--14, respectively. Blue filled-circles represent the peak brightness per bin in the dirty map, and their error bars show the $1$-$\sigma$ uncertainty on each measurement. The black line represents the pulse profile derived from the Effelsberg PSRIX single-dish data. The uncalibrated profiles from single-dish data are manually aligned (using an arbitrary scale) with the light curves obtained from dirty images, and are presented here only to show the consistency. The orange regions represent the bins used to search for the off-pulse emission in each pulsar.}
    \label{fig:light-curve}
\end{figure}

The interferometric data have been reduced in {\tt AIPS}\footnote{The Astronomical Image Processing System ({\tt AIPS}) is a software package produced and maintained by the National Radio Astronomy Observatory (NRAO).} \citep{greisen2003} and {\tt Difmap} \citep{shepherd1994} following standard procedures. A-priori amplitude calibration was performed using the known gain curves and system temperature measurements recorded on each station during the observations. For Jodrell Bank Mark2 in all observations, and Svetloe and Zelenchukskaya in the May 2017 observations, we used the nominal System Equivalent Flux Density (SEFD) values instead.
We removed the bad data (mainly affected by radio frequency interference) by manual flagging on each observation. We then fringe-fitted and bandpass calibrated the data using the fringe finders and phase calibrators. We imaged and self-calibrated the phase calibrator to improve the final calibration of the data. The obtained solutions were subsequently transferred to the pulsar data. We produced images for each individual bin to reveal the location of the pulsed emission as well as to search for any emission in the off-pulse region. For each bin, we measured the peak brightnesses at the position of the pulsars and the rms noise level of the image. We imaged the bins with the strongest pulsed emission and created a model source by fitting the $uv$ data. This model was used to self-calibrate the data, and the obtained solutions were applied to all bins. This step significantly improved the final results on the pulsars as we minimized the errors produced in the phase-referencing technique. We finally imaged the pulsed emission by combining the bins that clearly showed pulsed emission, and the off-pulse emission by selecting all bins away this region (see Fig.~\ref{fig:light-curve}).

We imaged the four epochs separately, and then we also combined the images from the two epochs for each pulsar to improve the sensitivity. Final images for the off-pulse emission were obtained by using a natural weighting \citep{briggs1995}, which also maximizes the reached sensitivity, on a field of up to a couple of arcsec centered on the pulsar positions. Different approaches were conducted during imaging to search for extended emission in the field: considering only the shortest baselines (with lengths $\lesssim 10~\mathrm{M\lambda}$; see Fig.~\ref{fig:uvcoverage} for reference), and $uv$-tapering the data to enhance the contribution of possible extended emission in the images.

\section{Results}\label{sec:results}

Both pulsars, \ba and \bb, were clearly detected in both the EVN and Effelsberg single-dish data during the pulsed emission.
We first used the obtained light-curves to determine the bins containing pulsed emission and the ones to be used for the off-pulse search (see Fig.~\ref{fig:light-curve}). In both pulsars, the pulsed emission is significantly detected along four bins.
No significant emission (above the rms noise level) is found in any of the other bins and no other sources are visible in the imaged field of view.

The four bins with significant pulsed emission were combined together to obtain a measurement of the flux density of the pulsed region at each epoch. We combined the bins outside this region (discarding also a few adjacent ones) to search for off-pulse emission with higher sensitivity.
No significant emission above the noise level was detected in the off-pulse region from either of the pulsars at any of the epochs, and even after combining the two epochs for each of the pulsars. Below we summarize the obtained results for both \ba and \bb separately.

\subsection{\ba}

Pulsed emission is clearly detected in bins 16--19 (see Fig.~\ref{fig:light-curve}, top, for reference). We combined these four bins to produce an image of the pulsed emission, and bins 1--11 and 24--64 to search for the off-pulse emission.
The pulsed emission is detected with a flux density of $110 \pm 30$ and $220 \pm 14~\mathrm{mJy}$ at the first and second epoch, respectively. The synthesized beams of these images are $30 \times 19$ and $23 \times 10~\mathrm{mas^2}$, respectively.

\begin{figure}[!t]
    \includegraphics[width=0.49\textwidth]{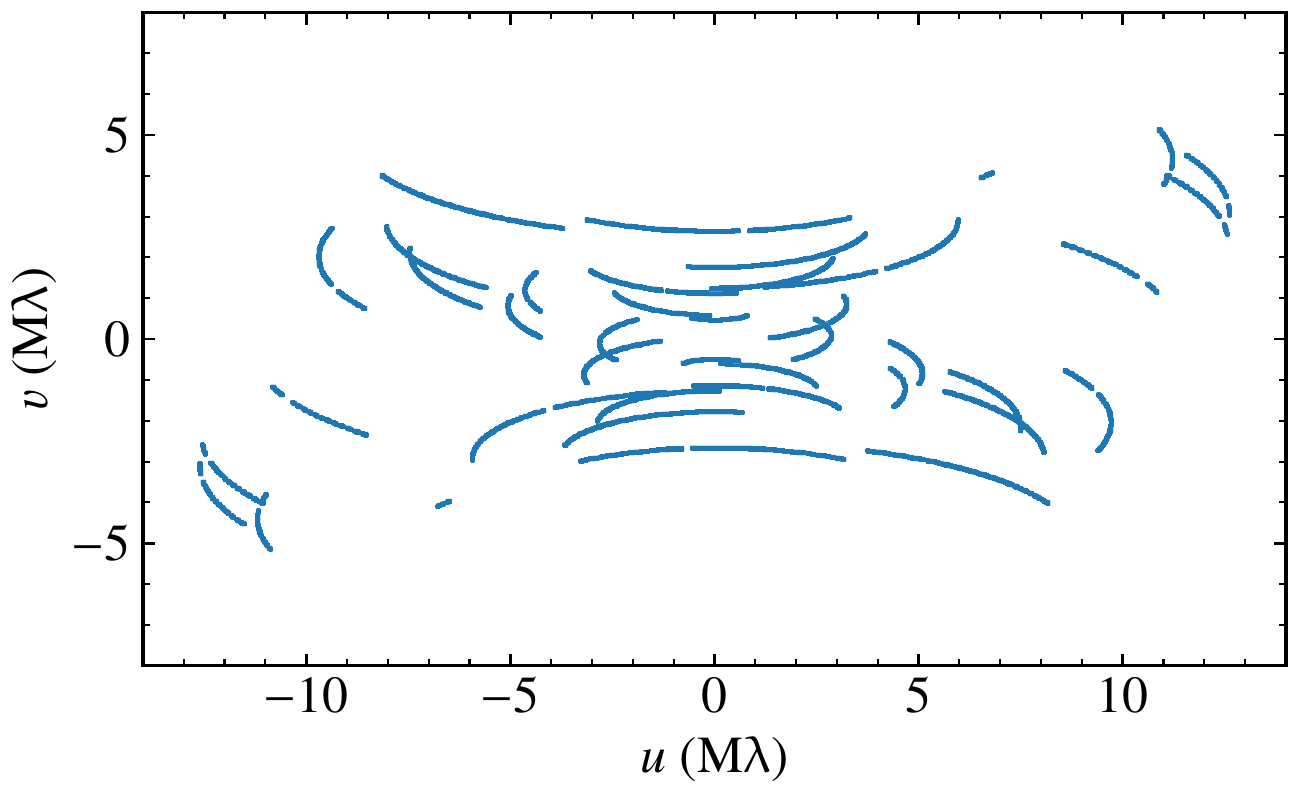}
    \caption{Resulting $uv$-coverage for the combined data of \ba used to search for the off-pulse emission. Only lengths $\lesssim 10~\mathrm{M\lambda}$ were considered to improve the sensitivity on the possible diffuse emission. The central frequency is 1.39~GHz with a bandwidth of 128~MHz. We note that given the resulting $uv$-coverage, our observations are only sensitive to angular scales $\lesssim 500~\mathrm{mas}$.}
    \label{fig:uvcoverage}
\end{figure}
\begin{figure*}[t]
    \begin{center}
        \includegraphics[width=0.9\textwidth]{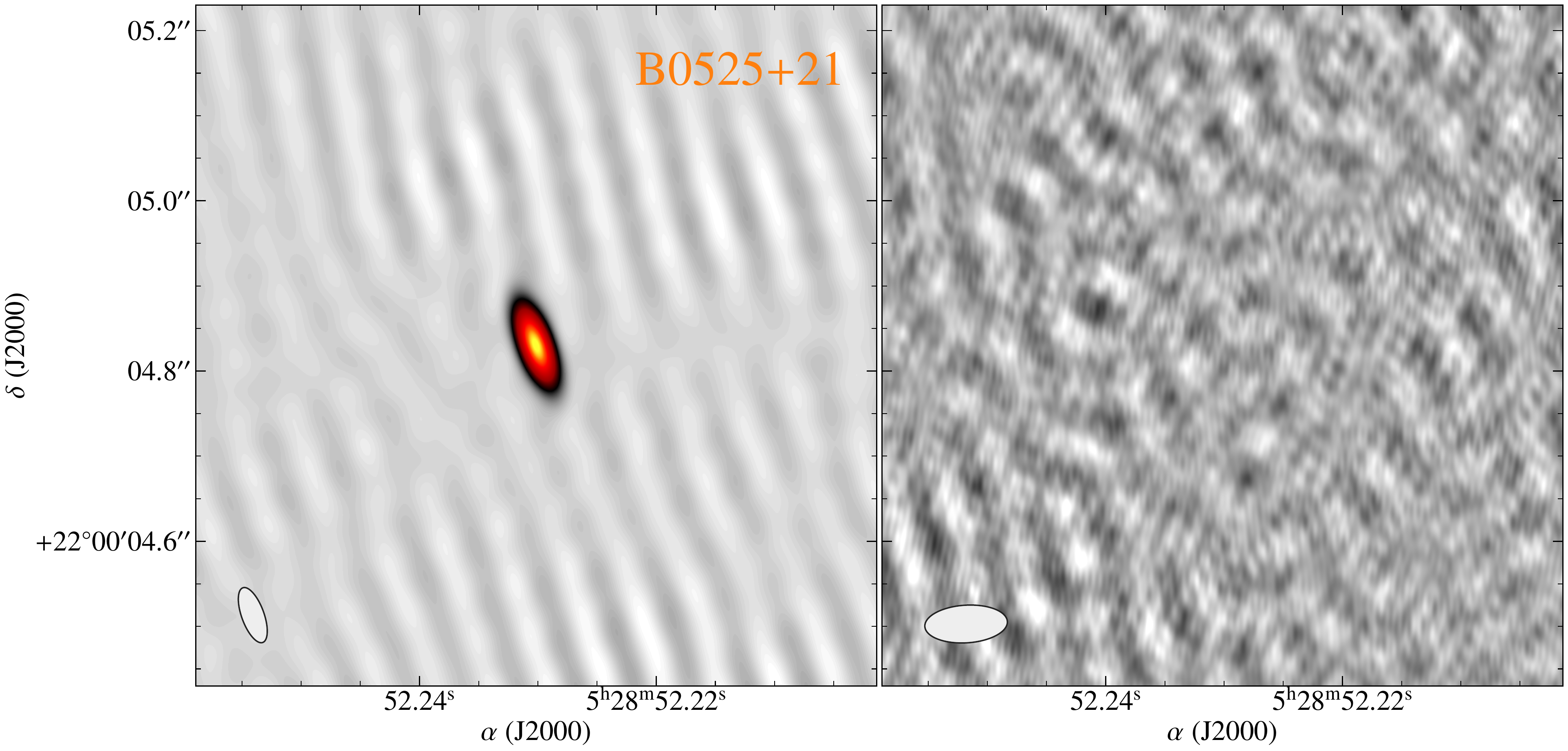}\\[+10pt]
        \includegraphics[width=0.9\textwidth]{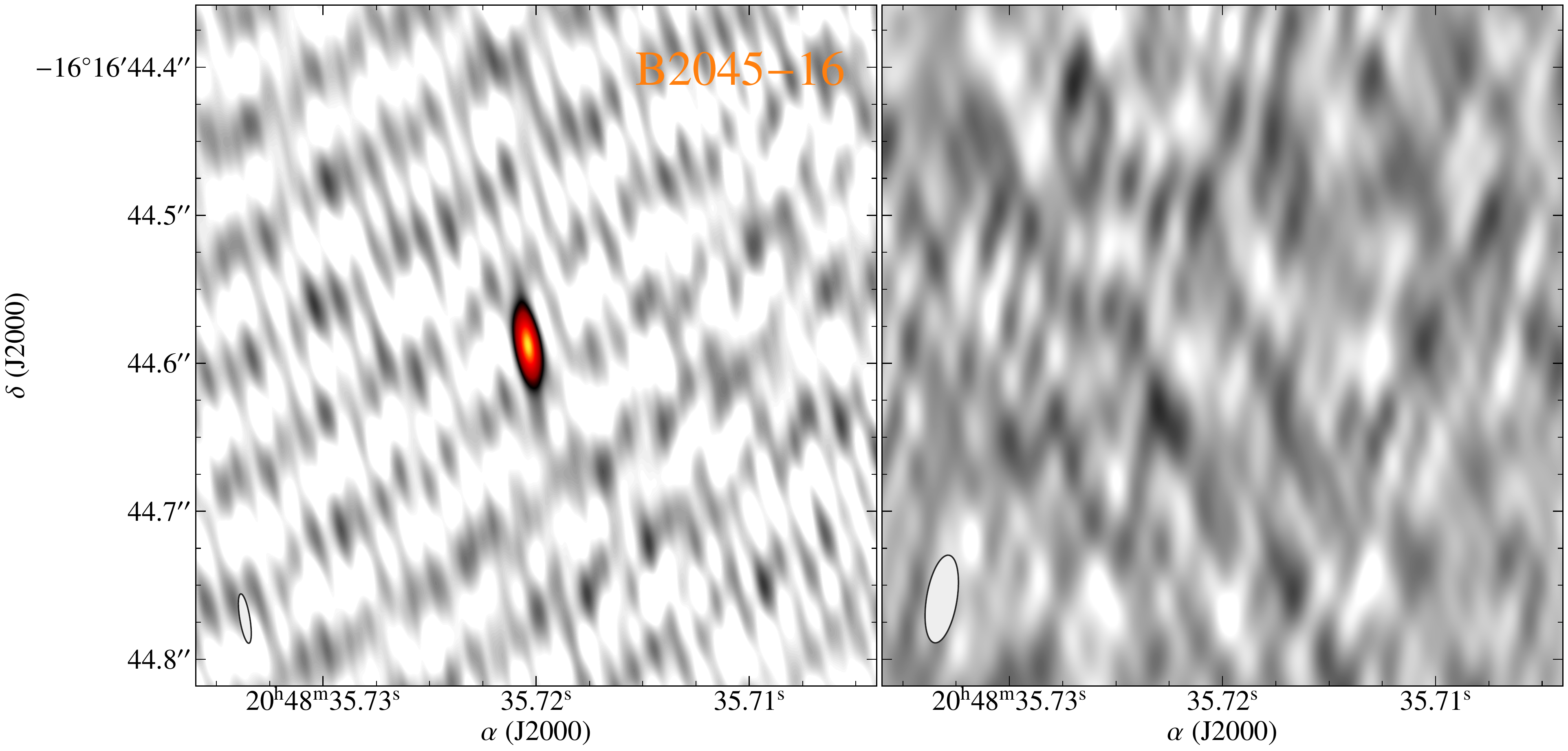}
    \end{center}
    \caption{EVN images of \ba (top) and \bb (bottom). Left panels show the images for the pulsed emission from each pulsar, whereas the right panels show those for the off-pulse regions (see text for further details). The color-scale starts above the 3-$\sigma$ rms noise level of each image, and gray-scale covers the range $-3$ to $3$-$\sigma$. Synthesized beams are shown at the bottom left corner of each panel. From left to right, and top to bottom, the rms noise values for each image are $2\ \mathrm{mJy\ beam^{-1}}$, $14\ \uJyb$, $3\ \mathrm{mJy\ beam^{-1}}$, and $32\ \uJyb$. The peak brightnesses of the pulsed emission are $110$ and $323\ \mathrm{mJy\ beam^{-1}}$ for \ba (top left) and \bb (bottom left).}
    \label{fig:images}
\end{figure*}

\begin{figure*}[t]
    \includegraphics[width=1\textwidth]{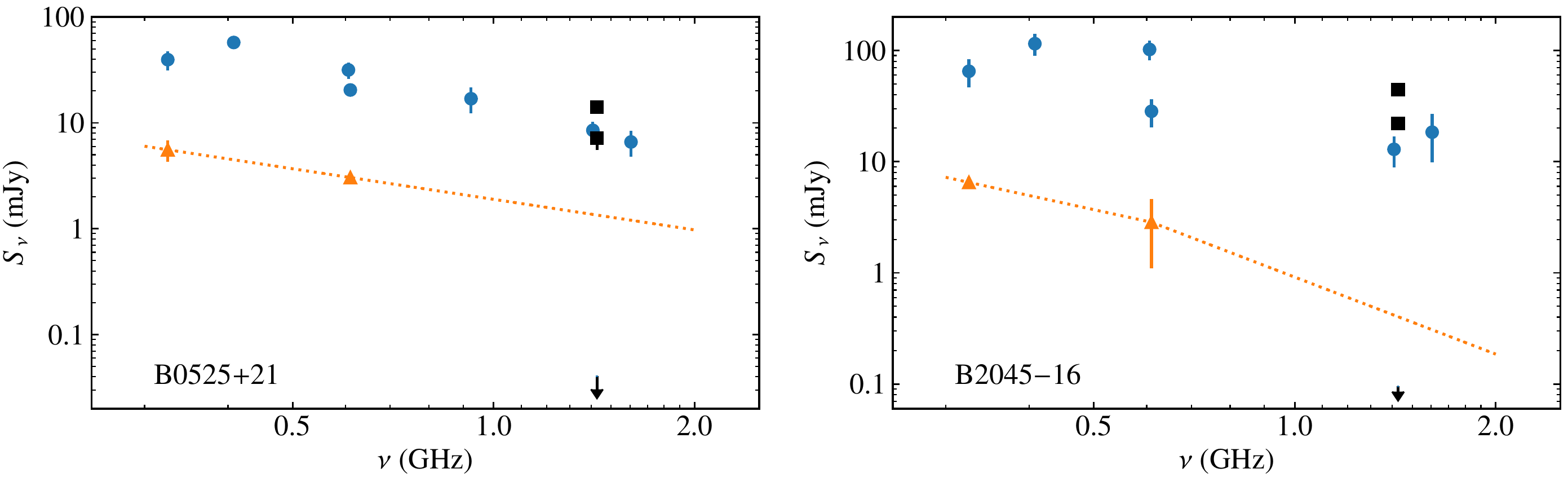}
    \caption{Combined spectra of \ba (left) and \bb (right) for the pulsed as well as off-pulse emission. Blue circles represent the pulsed emission reported by \citet{basu2012}, \citet{maron2000}, and \citet{lorimer1995}. The orange triangles represent the off-pulse emission reported by \citet{basu2012}. Black squares show the EVN results for the pulsed emission, and the black arrows represent the $3$-$\sigma$ upper-limits derived for the off-pulse emission from the EVN data. The dotted orange lines show the extrapolated spectrum for the off-pulse emission derived from \citet{basu2012} for reference. A spectral break has been included for \bb to show the worst-case scenario, see text for details.}
    \label{fig:spectrum}
\end{figure*}
No off-pulse emission is detected above the rms noise level --- neither at the individual epochs (with rms noise levels of $23$ and $17~\uJyb$) nor in the combined data (with a rms noise level of $14~\uJyb$). We thus constrained compact off-pulse emission to be $< 0.4\%$ of the pulsed emission averaged over the period at 3-$\sigma$ confidence level. No additional sources were detected in the field around the pulsar (up to a few arcseconds). To strengthen the possible extended emission, we produced images by applying a tapering in the $uv$-plane that drops the contribution of the most extended baselines and by only considering the shortest baselines directly. Null results were obtained in all cases with rms noise levels of $15$--$20~\uJyb$ and synthesized beams from $15 \times 30$ to $25 \times 38~\mathrm{mas^2}$.

\subsection{\bb}

In this case, the pulsed emission is detected in bins 11--14 (see Fig.~\ref{fig:light-curve}, bottom, for reference). We followed the same procedure as in \ba, producing an image for the pulsed emission by combining the mentioned bins. The resulting pulsed emission shows flux densities of $350 \pm 40$ and $590 \pm 30~\mathrm{mJy}$ at each epoch, with synthesized beams of $54 \times 26~\mathrm{mas^2}$ and $77 \times 18~\mathrm{mas^2}$, respectively.

Presence of off-pulse emission was searched by combining bins 1--7 and 18--64. No significant emission is reported above the rms noise levels of $66$ and $37~\uJyb$ at individual epochs, or above $32~\uJyb$ for the combined data. Compact off-pulse emission is thus constrained to be $< 0.3\%$ of the pulsed emission averaged over the period at 3-$\sigma$ confidence level.\\

Figure~\ref{fig:images} shows the final images obtained for both the pulsed and off-pulse emission for each pulsar.
Figure~\ref{fig:spectrum} shows the derived spectra for \ba and \bb combining the
flux density measurements for the pulsed emission (averaged over the period) as well as the off-pulse detections from \citet{basu2011} and the results reported in this work. We discuss the obtained results in the following section.

\section{Discussion}\label{sec:discussion}

The obtained results rule out the presence of off-pulse emission at 1.39~GHz originated from the magnetospheres of the pulsars \ba and \bb with peak brightnesses larger than $42$ and $96~\uJyb$ at $3$-$\sigma$ confidence level.

In the following subsections, we discuss the results obtained in this work in the context of \citet{basu2011,basu2012}'s results. We first discuss that our non-detections are unlikely to be explained by a spectral break or temporal variability. These conclusions lead to the possibility of an extended off-pulse emission, which could be originated outside the pulsar magnetosphere. We then discuss the possible origins of the earlier claimed off-pulse emission.

\subsection{Spectral break and temporal variability}

\citet{basu2012} suggested that the off-pulse and pulsed emission should have a similar radio spectra if both share a magnetospheric origin.
Considering the average flux densities reported by \citet{basu2011} for the off-pulse emission, one could expect a flux density emission of $\sim 1.2$ and $0.4\ \mathrm{mJy}$ at our observed frequency of 1.39~GHz for \ba and \bb, respectively (see Fig.~\ref{fig:spectrum}). We note that in the case of \bb, a frequency turnover is observed in the pulsed emission above $\sim 0.8~\mathrm{GHz}$ \citep{basu2012} and a similar turnover should be expected for the off-pulse emission (assuming the same magnetospheric origin). The obtained upper-limits from the EVN observations ($\lesssim 42$ and $96\ \uJyb$ at 3-$\sigma$ confidence level) are significantly below these expectations and exclude this hypothesis at $86$ and $13$-$\sigma$ levels.

The presence of a spectral break between 610~MHz and 1.39~GHz would thus be required to reconcile the off-pulse emission measurements. However, this approach would require post-break spectral indices of $\alpha \lesssim -5.2$ and $-3.2$ for \ba and \bb, respectively (where $\alpha$ is defined as $S_\nu \propto \nu^{\alpha}$).
A spectral break in the off-pulse emission itself would be unusual, especially when the spectral properties at lower frequencies are found to be similar to the pulsed emission. But in any case, the implied spectral indices are unrealistically steep \citep[see e.g.][]{bates2013}. Hence, we believe that a spectral break would not have caused the non-detections of the off-pulse emission.

Another possible reconciliation between all results could be the existence of variability.
The EVN and the GMRT observations are separated by several years. Refractive scintillations could potentially cause variability on such long timescales. Such a variability could typically change the observed flux densities by a factor of a few. A comparison of our pulsed flux density measurements with those from others (Figure~\ref{fig:spectrum}) also suggest that the changes would not have been more than a few. However, for both \ba and \bb, our 3-$\sigma$ upper limits are lower than the expected off-pulse flux densities by more than an order of magnitude and a factor of nearly four, respectively. Hence, our observations were sensitive enough to make a high significant detection, at least for \ba, even in the presence of an unfavorable refractive scintillation. Furthermore, the two pulsars are located in very different parts of the sky and are unlikely to exhibit correlated changes in flux density due to scintillation. Hence, consistent and sensitive non-detections from the EVN observations at two different epochs separated by more than three months for either of the pulsars, makes it unlikely that our results are affected by a temporal variability caused by refractive scintillations.

From the multiple measurements at lower frequencies \citep{basu2011,basu2012}, there is also no signature of any significant \emph{intrinsic} variability that could have caused our non-detections.

\subsection{Possible origin of the off-pulse emission}

\citet{basu2012} favored the magnetospheric origin to explain the off-pulse emission. However, under this scenario the emission should have still arised as a detectable compact object on the EVN scales.
\citet{basu2013} proposed a possible physical mechanism for the off-pulse emission: cyclotron resonance instability near the light cylinder.
The authors demonstrate the viability of their proposed emission mechanism and present a ``damping frequency'' which suggests the maximum frequency that can be emitted for the off-pulse emission to take place in the outer magnetosphere near the light cylinder. However, we note that the damping frequency for their assumed parameters is only about 240\,MHz for \ba, which is much lower than the highest frequency of 610\,MHz at which detection of off-pulse emission has been claimed. Moreover, the claimed off-pulse emission shows a trend similar to that of the on-pulse emission, and does not show any particular indications of damping till 610\,MHz. By assuming slightly more energetic secondary particles, the expected damping frequency can easily be near 500\,MHz and still satisfy the other constraints proposed by \citet[][e.g., eq.\ 12]{basu2013}. The fractional offset of the damping frequency from our observing frequency of 1.39\,GHz would then require similar arguments about the emission heights as those to justify the above mentioned discrepancy between damping and observing frequencies of 240 and 610\,MHz, respectively. We note that there are no such inconsistencies for \bb and off-pulse emission from this source at our observing frequency can easily be shown to be viable assuming particles with only slightly more energies (still within the permissible ranges.)
Thus even considering the frequency constraints from the above proposed emission mechanism, the non-detection of any off-pulse emission in our observations perhaps points to the possible existence of an extended structure, like a PWN. 

The shortest baselines from the EVN observations are $\sim 1\ \mathrm{M\lambda}$ (see Fig.~\ref{fig:uvcoverage}), which at the observed frequency (1.39~GHz) imply an angular scale of $\sim 500\ \mathrm{mas}$. Any emission larger than that value would thus be completely resolved out in the resulting images.
Given the distances to \ba and \bb, we then exclude physical sizes for the off-pulse emission of $\lesssim 610$ and $475\ \mathrm{au}$, respectively (or $0.0030$ and $0.0023\ \mathrm{pc}$).
Simultaneously, the GMRT observations \citep{basu2011,basu2012}, with much shorter baselines (and thus lower resolution), constrained the reported off-pulse emission to be $\lesssim 0.09$ and $0.04\ \mathrm{pc}$ for \ba and \bb, respectively \citep{basu2011}.

Combining all these results we constrain the putative PWNe to be $\sim (0.61$--$19) \times 10^3\ \mathrm{au}$ for \ba and $\sim (0.48$--$8.3) \times 10^3\ \mathrm{au}$ for \bb. We note that these values are significantly smaller than typical sizes of PWNe that range from $\sim 20$ to $200 \times 10^3\ \mathrm{au}$ \citep[$\sim 0.1$ to $1\ \mathrm{pc}$,][]{gaensler2006}.
However, as discussed in \citet{basu2011}, whereas a static PWN can be ruled out, the case of a relatively compact bow shock PWN (where the bow shock is generated where the pulsar wind is confronted by the ram pressure of the pulsar's motion through the ISM) could explain the observed emission. Following equations (2--6) in \citet{basu2011}, the obtained size limits for the off-pulse emission require ISM densities of $1.1 \times 10^{-3}$--$1.2 \times 10^{-6}\ \mathrm{cm^{-3}}$ for \ba and $6.9 \times 10^{-4}$--$2.3 \times 10^{-6}\ \mathrm{cm^{-3}}$ for \bb.
The upper limits on these ranges are slightly lower but still consistent with typical ISM densities \citep{basu2012}. The presence of bow shock PWNe thus still seems plausible. With our EVN observations, we cannot confirm or rule-out the presence of above potential compact PWNe. However, radio observations covering intermediate angular scales (i.e., a few hundred mas to a few arcseconds) are planned which will help in clarifying this scenario.

Finally, we also note that \citet{basu2011} conducted several tests to rule out any contamination from instrumental leakages which might appear as off-pulse emission. However, \citet{Ravi2018} have recently indicated towards another possibility of a spectral leakage which might in fact appear as off-pulse emission for \textit{dispersed pulsed signals}. Our above mentioned future observations would also help in concluding on that front.

\section{Conclusions} \label{sec:conclusions}

\ba and \bb were reported to exhibit off-pulse radio emission originated from the respective pulsar's magnetosphere. We report the first VLBI observations of these two pulsars to unveil any compact off-pulse emission and understand its nature.
The obtained results are inconsistent with the presence of a \textit{magnetospherically} originated emission in the off-pulse region. Spectral break or variability are also ruled out as possible explanations, given the stringent and consistent upper limits from multiple-epoch observations of both the pulsars. However, the results are still compatible with the presence of compact, bow shock PWNe around the pulsars with sizes of $\sim 10^{3\text{--}4}\ \mathrm{au}$.
Future observations sensitive to intermediate angular scales and frequencies between those of the GMRT and the EVN observations have been planned to probe the existence of such extended and diffuse PWNe around \ba and \bb.

\begin{acknowledgements}

The European VLBI Network is a joint facility of independent European, African, Asian, and North American radio astronomy institutes. Scientific results from data presented in this publication are derived from the following EVN project code: EM127. We thank the staff of the Effelsberg Radio Telescope, and in particular R.~Karuppusam and U.~Bach, for their support with simultaneous pulsar recording.
BM acknowledges support from the Spanish Ministerio de Econom\'ia y Competitividad (MINECO) under grants AYA2016-76012-C3-1-P and MDM-2014-0369 of ICCUB (Unidad de Excelencia ``Mar\'ia de Maeztu'').
YM acknowledges use of the funding from the European Research Council under the European Union's Seventh Framework Programme (FP/2007-2013)/ERC Grant Agreement no. 617199.
This research made use of APLpy, an open-source plotting package for Python hosted at \url{http://aplpy.github.com}, Astropy, a community-developed core Python package for Astronomy \citep{astropy2013}, and Matplotlib \citep{hunter2007}.

\end{acknowledgements}

\bibliographystyle{aa}
\bibliography{bibliography.bib}

\begin{thebibliography}{22}
\expandafter\ifx\csname natexlab\endcsname\relax\def\natexlab#1{#1}\fi

\bibitem[{{Astropy Collaboration} {et~al.}(2013){Astropy Collaboration},
  {Robitaille}, {Tollerud}, {Greenfield}, {Droettboom}, {Bray}, {Aldcroft},
  {Davis}, {Ginsburg}, {Price-Whelan}, {Kerzendorf}, {Conley}, {Crighton},
  {Barbary}, {Muna}, {Ferguson}, {Grollier}, {Parikh}, {Nair}, {Unther},
  {Deil}, {Woillez}, {Conseil}, {Kramer}, {Turner}, {Singer}, {Fox}, {Weaver},
  {Zabalza}, {Edwards}, {Azalee Bostroem}, {Burke}, {Casey}, {Crawford},
  {Dencheva}, {Ely}, {Jenness}, {Labrie}, {Lim}, {Pierfederici}, {Pontzen},
  {Ptak}, {Refsdal}, {Servillat}, \& {Streicher}}]{astropy2013}
{Astropy Collaboration}, {Robitaille}, T.~P., {Tollerud}, E.~J., {et~al.} 2013,
  \aap, 558, A33

\bibitem[{{Basu} {et~al.}(2011){Basu}, {Athreya}, \& {Mitra}}]{basu2011}
{Basu}, R., {Athreya}, R., \& {Mitra}, D. 2011, \apj, 728, 157

\bibitem[{{Basu} {et~al.}(2012){Basu}, {Mitra}, \& {Athreya}}]{basu2012}
{Basu}, R., {Mitra}, D., \& {Athreya}, R. 2012, \apj, 758, 91

\bibitem[{{Basu} {et~al.}(2013){Basu}, {Mitra}, \& {Melikidze}}]{basu2013}
{Basu}, R., {Mitra}, D., \& {Melikidze}, G.~I. 2013, \apj, 772, 86

\bibitem[{{Bates} {et~al.}(2013){Bates}, {Lorimer}, \& {Verbiest}}]{bates2013}
{Bates}, S.~D., {Lorimer}, D.~R., \& {Verbiest}, J.~P.~W. 2013, \mnras, 431,
  1352

\bibitem[{{Blandford} {et~al.}(1973){Blandford}, {Ostriker}, {Pacini}, \&
  {Rees}}]{Blandford73}
{Blandford}, R.~D., {Ostriker}, J.~P., {Pacini}, F., \& {Rees}, M.~J. 1973,
  \aap, 23, 145

\bibitem[{{Briggs}(1995)}]{briggs1995}
{Briggs}, D.~S. 1995, in Bulletin of the American Astronomical Society,
  Vol.~27, American Astronomical Society Meeting Abstracts, 1444

\bibitem[{{Cohen} {et~al.}(1983){Cohen}, {Cotton}, {Geldzahler}, \&
  {Marcaide}}]{cohen1983}
{Cohen}, N.~L., {Cotton}, W.~D., {Geldzahler}, B.~J., \& {Marcaide}, J.~M.
  1983, \apj, 264, 273

\bibitem[{{Gaensler} \& {Slane}(2006)}]{gaensler2006}
{Gaensler}, B.~M. \& {Slane}, P.~O. 2006, \araa, 44, 17

\bibitem[{{Greisen}(2003)}]{greisen2003}
{Greisen}, E.~W. 2003, in Astrophysics and Space Science Library, Vol. 285,
  Information Handling in Astronomy - Historical Vistas, ed. A.~{Heck}, 109

\bibitem[{{Hankins} {et~al.}(1993){Hankins}, {Moffett}, {Novikov}, \&
  {Popov}}]{hankins1993}
{Hankins}, T.~H., {Moffett}, D.~A., {Novikov}, A., \& {Popov}, M. 1993, \apj,
  417, 735

\bibitem[{Hunter(2007)}]{hunter2007}
Hunter, J.~D. 2007, Computing In Science \& Engineering, 9, 90

\bibitem[{{Keimpema} {et~al.}(2015){Keimpema}, {Kettenis}, {Pogrebenko},
  {Campbell}, {Cim{\'o}}, {Duev}, {Eldering}, {Kruithof}, {van Langevelde},
  {Marchal}, {Molera Calv{\'e}s}, {Ozdemir}, {Paragi}, {Pidopryhora},
  {Szomoru}, \& {Yang}}]{keimpema2015}
{Keimpema}, A., {Kettenis}, M.~M., {Pogrebenko}, S.~V., {et~al.} 2015,
  Experimental Astronomy, 39, 259

\bibitem[{{Kettenis} \& {Keimpema}(2014)}]{kettenis2014}
{Kettenis}, M. \& {Keimpema}, A. 2014, in Proceedings of the 12th European VLBI
  Network Symposium and Users Meeting (EVN 2014). 7-10 October 2014. Cagliari,
  88

\bibitem[{{Lazarus} {et~al.}(2016){Lazarus}, {Karuppusamy}, {Graikou},
  {Caballero}, {Champion}, {Lee}, {Verbiest}, \& {Kramer}}]{lazarus2016}
{Lazarus}, P., {Karuppusamy}, R., {Graikou}, E., {et~al.} 2016, \mnras, 458,
  868

\bibitem[{{Lorimer} {et~al.}(1995){Lorimer}, {Yates}, {Lyne}, \&
  {Gould}}]{lorimer1995}
{Lorimer}, D.~R., {Yates}, J.~A., {Lyne}, A.~G., \& {Gould}, D.~M. 1995,
  \mnras, 273, 411

\bibitem[{{Maron} {et~al.}(2000){Maron}, {Kijak}, {Kramer}, \&
  {Wielebinski}}]{maron2000}
{Maron}, O., {Kijak}, J., {Kramer}, M., \& {Wielebinski}, R. 2000, \aaps, 147,
  195

\bibitem[{{Perry} \& {Lyne}(1985)}]{perry1985}
{Perry}, T.~E. \& {Lyne}, A.~G. 1985, \mnras, 212, 489

\bibitem[{{Ravi} \& {Deshpande}(2018)}]{Ravi2018}
{Ravi}, K. \& {Deshpande}, A.~A. 2018, \apj, 859, 22

\bibitem[{{Sch{\"o}nhardt}(1973)}]{schonhardt1973}
{Sch{\"o}nhardt}, R.~E. 1973, Nature Physical Science, 243, 62

\bibitem[{{Shepherd} {et~al.}(1994){Shepherd}, {Pearson}, \&
  {Taylor}}]{shepherd1994}
{Shepherd}, M.~C., {Pearson}, T.~J., \& {Taylor}, G.~B. 1994, in Bulletin of
  the American Astronomical Society, Vol.~26, Bulletin of the American
  Astronomical Society, 987

\bibitem[{{Strom} \& {van Someren Greve}(1990)}]{strom1990}
{Strom}, R.~G. \& {van Someren Greve}, H.~W. 1990, \apss, 171, 351

\end{thebibliography}

\end{document}